\documentclass[
reprint, amsmath,amssymb,aps,twocolumn]{revtex4-2}
\usepackage[force]{feynmp-auto}
\usepackage{subfigure}
\usepackage{graphicx}
\usepackage{float}
\usepackage{dcolumn}
\usepackage{bm}
\usepackage{physics}
\usepackage[backref]{hyperref}

\begin{document}
	\preprint{APS/123-QED}
	\title{The density-functional theory of quantum droplets}
	\author{Fan Zhang}
	\affiliation{School of Physics, Peking University, Beijing 100871, China}
	\affiliation{CAS Center For Excellence in Quantum Information and Quantum Physics, Hefei 230026, China}
	\author{Lan Yin}
	\email{yinlan@pku.edu.cn}
	\affiliation{School of Physics, Peking University, Beijing 100871, China}
	\date{\today}
	\begin{abstract}
		In quantum droplets, the mean-field energy is comparable to the Lee-Huang-Yang (LHY) energy.   In the Bogoliubov theory, the LHY energy of the quantum droplet has an imaginary part, but it is neglected for practical purposes.   So far, most theoretical studies of quantum droplets have been based on the extended Gross-Pitaevskii (GP) equation obtained by adding the LHY energy to the GP equation.  In this article, we present the density-functional theory of quantum droplets.  In our approach, the quantum fluctuations in quantum droplets, as described by an effective action, generate the correlation energy which is real and can be determined self-consistently. Using the density-functional theory, we calculate higher-order corrections to the energy, the quantum depletion fraction, and the excitations of the droplet. Our results for the ground-state energy and the quantum depletion fraction are compared with the Monte Carlo results and good agreement is found.  The implications of our theory are discussed.
	\end{abstract}
	\maketitle
	\section{Introduction}
The creation of quantum droplets has been a breakthrough in the research on ultracold atoms in recent years.  So far quantum droplets have been realized in various systems, e. g. in a dipolar Bose gas such as $^{164}\rm{Dy}$\cite{Kadau2016b,Ferrier-Barbut2016,Ferrier-Barbut2016a,Schmitt2016a,Wenzel2017} and $ ^{166}\rm{Er}$\cite{chomaz2016quantum}, and in binary boson mixture such as homonuclear $ ^{39}\rm{K}$ \cite{Cabrera2018,Cheiney2018,semeghini2018self} and heteronuclear  $ ^{39}\rm{K}$-$ ^{87}\rm{Rb}$ mixtures\cite{d2019observation}.  In these experiments, quantum droplets are generated by tuning the $s$-wave interaction by the Feshbach-resonance technique.  In these quantum droplets, the mean-field energy is tuned into a weakly-attractive energy, and the repulsive Lee-Huang-Yang (LHY) energy \cite{Lee1957} from quantum fluctuations becomes equally important.  The competition between these two energies results in the quantum-droplet state which is self-bound and stable.

Due to the mean-field instability, in the Bogoliubov theory which describes the Gaussian fluctuations around the uniform condensate, there are imaginary excitation energies in the long wavelength limit, implying the dynamical instability contradicting the experiments.  Petrov \cite{Petrov2015} pointed out that these unstable excitations have little contribution to the LHY energy and may be stable after renormalization by integrating out high-energy excitations.  In practice, the LHY energy with its imaginary part neglected is put into the Gross-Pitaevskii (GP) equation, which is the so-called extended Gross-Pitaevskii equation (EGPE)\cite{Petrov2015,wachtler2016ground} widely used in simulating quantum droplets.  The dynamic instability in the Bogoliubov theory is artificial as found in recent studies by the Beliaev theory \cite{gu2020phonon,2021xiong,zhang2022phonon}.  The phonon energy is stable after the higher-order quantum fluctuations are taken into account, for both the nondipolar Bose mixture \cite{gu2020phonon,2021xiong} and the single-component dipolar Bose gas \cite{zhang2022phonon}.  In this work, we present the density-functional theory of quantum droplets, which treats the quantum fluctuations self-consistently without the suffer of imaginary energies.  The essence of this method is including the effect of higher order fluctuations by renormalizing the $s$-wave coupling constants.  For the binary boson mixture, the ground state energy obtained in our approach are in better agreement with the diffusion Monte Carlo (DMC) simulation result \cite{cikojevic2019universality} than the EGPE result \cite{Petrov2015}.  For the dipolar quantum droplet, our results about the quantum depletion fit the quantum Monte  Carlo (QMC) results \cite{bottcher2019dilute} better than the Bogoliubov theory.  The implications of our theory are discussed.

	\section{density-functional theory}
	
	We study a multi-component Bose gas with its Hamiltonian given by
\begin{gather}
	H=\int  d \vb r \sum_{\sigma} \Big[ \psi^\dagger_{\sigma} (\vb r)\big(- \frac{\hbar^2\nabla^{2}}{2m_{\sigma}} +V_\sigma(\vb r) \big) \psi_{\sigma}(\vb r) \\ \notag
	+\frac{1}{2}\int d \vb r' \sum_{{\sigma}'} U_{{\sigma} {\sigma}'}(\vb r-\vb r')\psi^\dagger_{\sigma}(\vb r) \psi^\dagger_{\sigma'}(\vb r') \psi_{\sigma'}(\vb r')\psi_{\sigma}(\vb r)\Big],
\end{gather}
	where $\psi_{\sigma}(\vb r)$ is the boson-field operator for the $\sigma$-component, $m_\sigma$ is the mass, $V_\sigma(\vb r)$ is the trap potential, and $ U_{{\sigma} {\sigma}'}(\vb r-\vb r')$ is the interaction between bosons.  For a BEC ground state, the condensate wavefunction is given by $\psi_{0\sigma}(\vb r)=\langle \psi_{\sigma}(\vb r) \rangle=\sqrt{n_{0\sigma}(\vb r)}e^{i\phi_{\sigma}(\vb r)}  $ where  $n_{0\sigma}(\vb r)$ is the condensate density, and $\phi_{\sigma}(\vb r)$ is the condensate phase.  
	
	We consider the case that the temporal and spatial scales of variances are much larger than the intrinsic scales of the system.  The local equilibrium assumption (LEA) can be applied and the system is described by the effective action \cite{zhang2023hydrodynamics} 
	\begin{gather} \label{EA}
		S_{Eff}= \int d t \int d \vb r \big\{ -\sum_{\sigma} n_{\sigma}(\vb r)[\hbar \partial_t\phi_{\sigma }(\vb r)+V_{\sigma}(\vb r) \\ \notag
		+\frac{\hbar^2}{2m_\sigma}(\frac{|\nabla n_{\sigma}|^2(\vb r)}{4n_{\sigma}(\vb r)}+|\nabla \phi_{\sigma}(\vb r)|^2)]-\mathcal{E}_I(\vb r)\big\},  
	\end{gather}
	where $n_{\sigma}(\vb r)$ is the superfluid density, and $\mathcal{E}_I$ is the interaction-energy density of the uniform steady state with the lowest energy with the superfluid density $n_{\sigma}$ and phase gradient $\nabla\phi_{\sigma}$. In the effective action $S_{eff}$, there is no term containing the time derivative of the density \cite{ueda2010fundamentals}.  From the effective action $S_{eff}$, the superfluid hydrodynamic equations can be obtained \cite{zhang2023hydrodynamics}.  For the ground state, the superfluid phase is uniform, $\phi_{\sigma}$=0, for simplicity, the interaction energy density $\mathcal{E}_I$ can be separated into two parts, $\mathcal{E}_I=\mathcal{E}_{MF}+\mathcal{E}_C$, where the mean-field energy density is given by  
	\begin{equation}
		\mathcal{E}_{MF}(\vb r)=\frac{1}{2}\int d \vb r' \sum_{\sigma,{\sigma}'} U_{{\sigma} {\sigma}'}(\vb r-\vb r')n_{\sigma}(\vb r) n_{\sigma'} (\vb r').
	\end{equation}
	The correlation-energy density $\mathcal{E}_C$ as a function of densities comes from quantum fluctuations beyond the mean-field and shall be determined self-consistently as explained in the latter part of this section.  The ground-state energy density $\mathcal{E}$ is given by $\mathcal{E}=\mathcal{E}_K+\mathcal{E}_{MF}+\mathcal{E}_C$, where the kinetic plus potential energy density is given by 
	\begin{equation}
		\mathcal{E}_K(\vb r)=  \sum_{\sigma}n_{\sigma}(\vb r) \big(\frac{\hbar^2}{2m_{\sigma}}\frac{|\nabla n_{\sigma}|^2(\vb r)}{4n_{\sigma}(\vb r)} +V_\sigma(\vb r) \big).
	\end{equation}
	The superfluid-density distribution should satisfy the minimization condition of  the ground-state energy $E$,
\begin{gather}\label{dd}
	- \frac{\hbar^2\nabla^{2}\sqrt{n_{\sigma}(\vb r)}}{2m_{\sigma}\sqrt{n_{\sigma}(\vb r)}} +V_\sigma(\vb r) \\ \notag+\int  d \vb r' \sum_{{\sigma}'} U_{{\sigma} {\sigma}'}(\vb r-\vb r') n_{\sigma'}(\vb r')+ \frac{\partial \mathcal{E}_C}{\partial n_\sigma}=\mu_{\sigma},
\end{gather}
	where $\mu_\sigma=\partial E/\partial N_\sigma$ is the chemical potential, and $N_\sigma$ is the boson number of the $\sigma$-component. From Eq. (\ref{dd}), the density distribution of a nonuniform system can be solved, similar to solving EGPE.
	
	The central task is to obtain the correlation-energy density $\mathcal{E}_C$.  In the dilute region, the typical treatment is to consider the Gaussian fluctuations around the condensate and study the Bogoliubvov Hamiltonian.  However, for systems such as quantum droplets with crucial beyond-mean-field effects, such treatment is inadequate, the effects of higher-order fluctuations must be taken into account \cite{gu2020phonon,2021xiong,zhang2022phonon}.  Here we propose that these important fluctuation effects can be captured by the Gaussian fluctuations in the effective action Eq. (\ref{EA}), as given by
	\begin{widetext}
		\begin{gather}\label{fac}
			S_2=-\int d t \int d \vb r \sum_{\sigma} \big\{ \delta n_{\sigma}(\vb r)\hbar \partial_t \delta \phi_{\sigma }(\vb r)
			+\frac{\hbar^2}{2m_\sigma}(|\nabla\frac{ \delta n_{\sigma}(\vb r)}{2n_{\sigma}(\vb r)}|^2+|\nabla \delta \phi_{\sigma}(\vb r)|^2) \\ \notag
			+\frac{1}{2}\int d \vb r' \sum_{{\sigma}'} U_{{\sigma} {\sigma}'}(\vb r-\vb r')\delta n_{\sigma}(\vb r) \delta n_{\sigma'}(\vb r') + \frac{1}{2}\sum_{{\sigma}'} \chi_{{\sigma} {\sigma}'}(\vb r)\delta n_{\sigma}(\vb r) \delta n_{\sigma'}(\vb r) \big\}, 
		\end{gather}
	\end{widetext}
	where 
	\begin{equation}\label{chi}
		\chi_{{\sigma} {\sigma}'}=\frac{\partial^2 \mathcal{E}_C}{\partial n_{\sigma}\partial n_{\sigma'}}.
	\end{equation}
	The difference from the Bogoliubov theory is that in Eq. (\ref{fac}) the Gaussian fluctuations are density and phase fluctuations in the effective action, rather than the fluctuations around the condensate.  In this way, the crucial higher-order effects beyond the Bogoliubov theory are taken into account.  This renormalization to the Bogoliubov theory is equivalent to a local correction to the $s$-wave coupling constant given by $\chi$ in Eq. (\ref{fac}).  For a uniform system, the correlation energy can be obtained by integrating out the fluctuating fields $ \delta n_{\sigma}$ and  $\delta \phi_{\sigma}$ in the action $S_2$, and thus determined self-consistently.  Beyond the dilute region, higher-order fluctuations in the effective action should be also considered, and in principle, the correlation-energy density can still be obtained by integrating out all the fluctuating fields.
	
	\section{implications on quantum droplets}
	Although the quantum droplets are in the dilute region, the mean-field energy is very small and comparable to the LHY energy.  It is important to determine the correlation energy properly.  In the following, we study the two types of quantum droplets found in experiments, i.e. the binary boson mixture and dipolar Bose gas.
	
	\subsubsection{Binary boson mixture}
	For a uniform binary boson mixture with short-ranged interactions, the action $S_2$ in Eq. (\ref{fac}) describing the Gaussian fluctuations is equivalent to a renormalized Bogoliubov Hamiltonian given by
	\begin{gather}
		H_{Eff}=\sum_{\vb k, \sigma} \epsilon_{k} a^\dagger_{\vb k \sigma} a_{\vb k \sigma} \\ \notag
		+\sum_{\vb k, \sigma,\sigma'}g'_{{\sigma} {\sigma}'}\sqrt{n_{\sigma} n_{\sigma'}} [a^\dagger_{\vb k \sigma} a_{\vb k \sigma'}+\frac{1}{2} ( a_{\vb k \sigma} a_{-\vb k \sigma'}+H. C. ) ],
	\end{gather}
	where  $\epsilon_{k}=\hbar^2k^2/2m$, $a_{\vb k \sigma}$ is the boson annihlation operator, $g'_{{\sigma} {\sigma}'}=g_{{\sigma} {\sigma}'}+\chi_{{\sigma} {\sigma}'}$, and $g_{{\sigma} {\sigma}'}$ is the coupling constant between $\sigma$- and $\sigma'$-components.  Two types of excitations can be obtained from this Hamiltonian, i.e. phonon and magnon, and both energies are linearly dispersed in the long-wavelength limit.  The phonon speed $c_-$ and the magnon speed $c_+$ are given by 
	\begin{gather}
		c_\pm=\frac{1}{2\sqrt{m}}\{g'_{11}n_1+g'_{22}n_2\\\notag \pm \sqrt{(g'_{11}n_1-g'_{22}n_2)^2+4g_{12}^{'2} n_1 n_2}\}^{1/2}.
	\end{gather}
	In this case, the correlation energy is the renormalized LHY energy given by
	\begin{equation}\label{bce}
		\mathcal{E}_C=\frac{8m^4}{15\pi^2\hbar^3}(c_+^5+c_-^5).
	\end{equation}
	The speeds $c_\pm$  can be solved from Eq. (\ref{bce}) and (\ref{chi}).  For the dilute binary quantum droplet, $c_-\gg c_+$ and $|g_{{\sigma} {\sigma}'}| \gg |\chi_{{\sigma} {\sigma}'}|$, to the leading order, the correlation energy is given by the LHY energy in Ref. \cite{Petrov2015},
	\begin{gather}\label{ec1}
		\mathcal{E}_C \approx \frac{\sqrt{2m^3}}{15\pi^2\hbar^3}(g_{11}n_1+g_{22}n_2\\\notag- \sqrt{(g_{11}n_1-g_{22}n_2)^2+4g_{12}^2 n_1 n_2})^{5/2}.
	\end{gather}
	Its second derivatives $\chi_{{\sigma} {\sigma}'}$ can be computed and a positive phonon speed can be obtained in agreement with Ref. \cite{gu2020phonon,2021xiong}. 
	
\begin{figure*}
		\centering
		\includegraphics[scale=0.13]{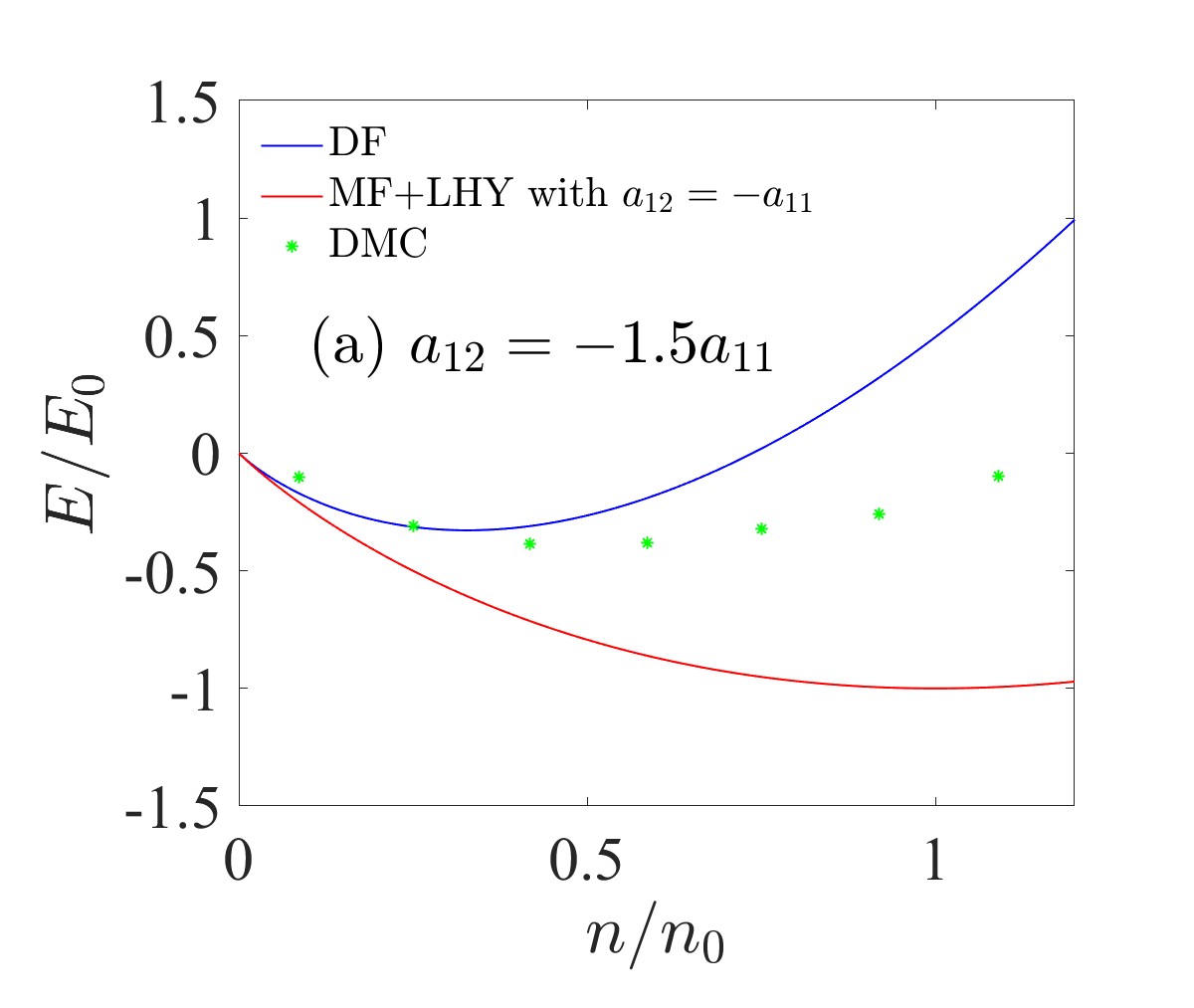}
		\hspace{0.1in}
		\includegraphics[scale=0.13]{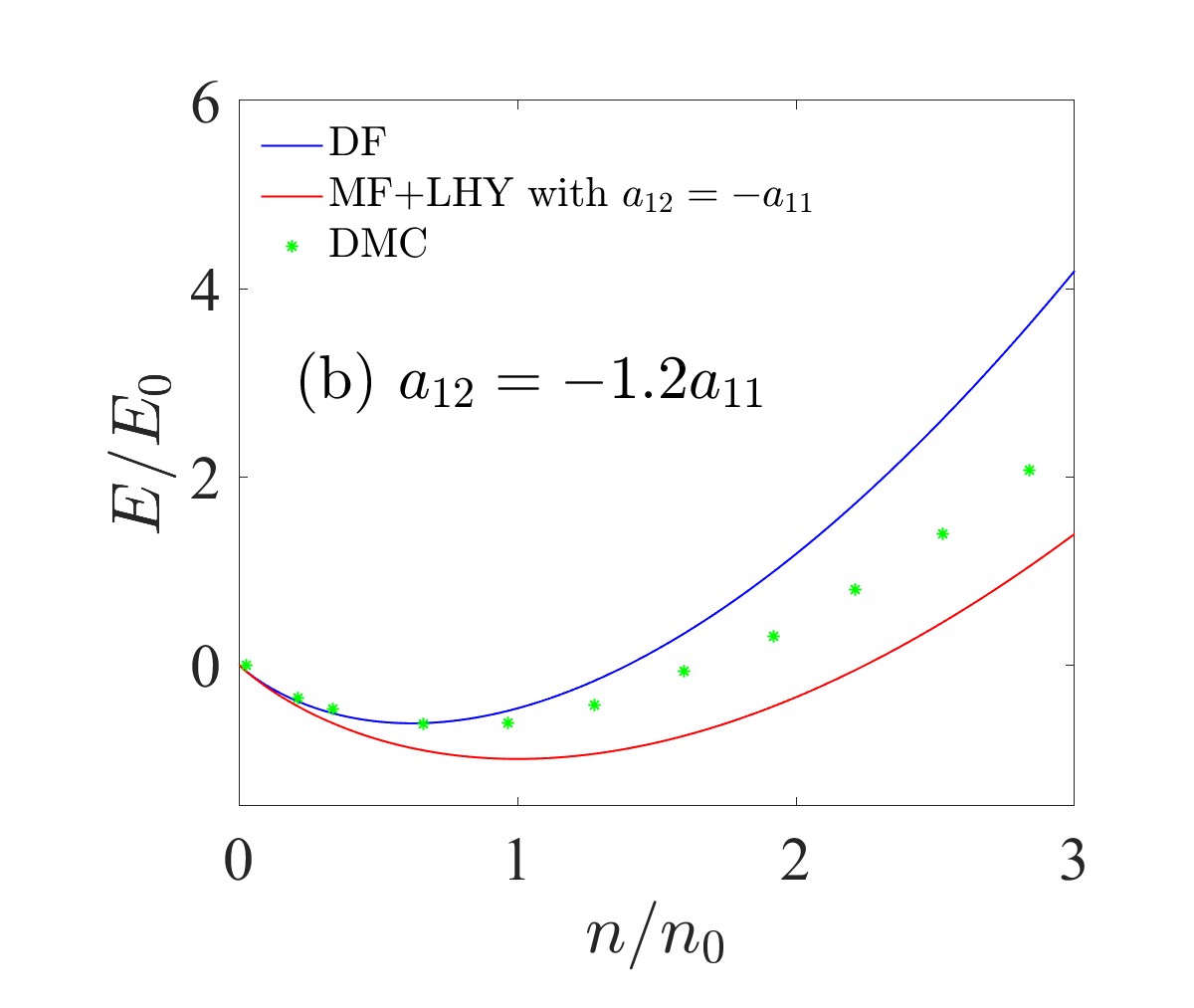}
		\hspace{0.1in}
		\includegraphics[scale=0.13]{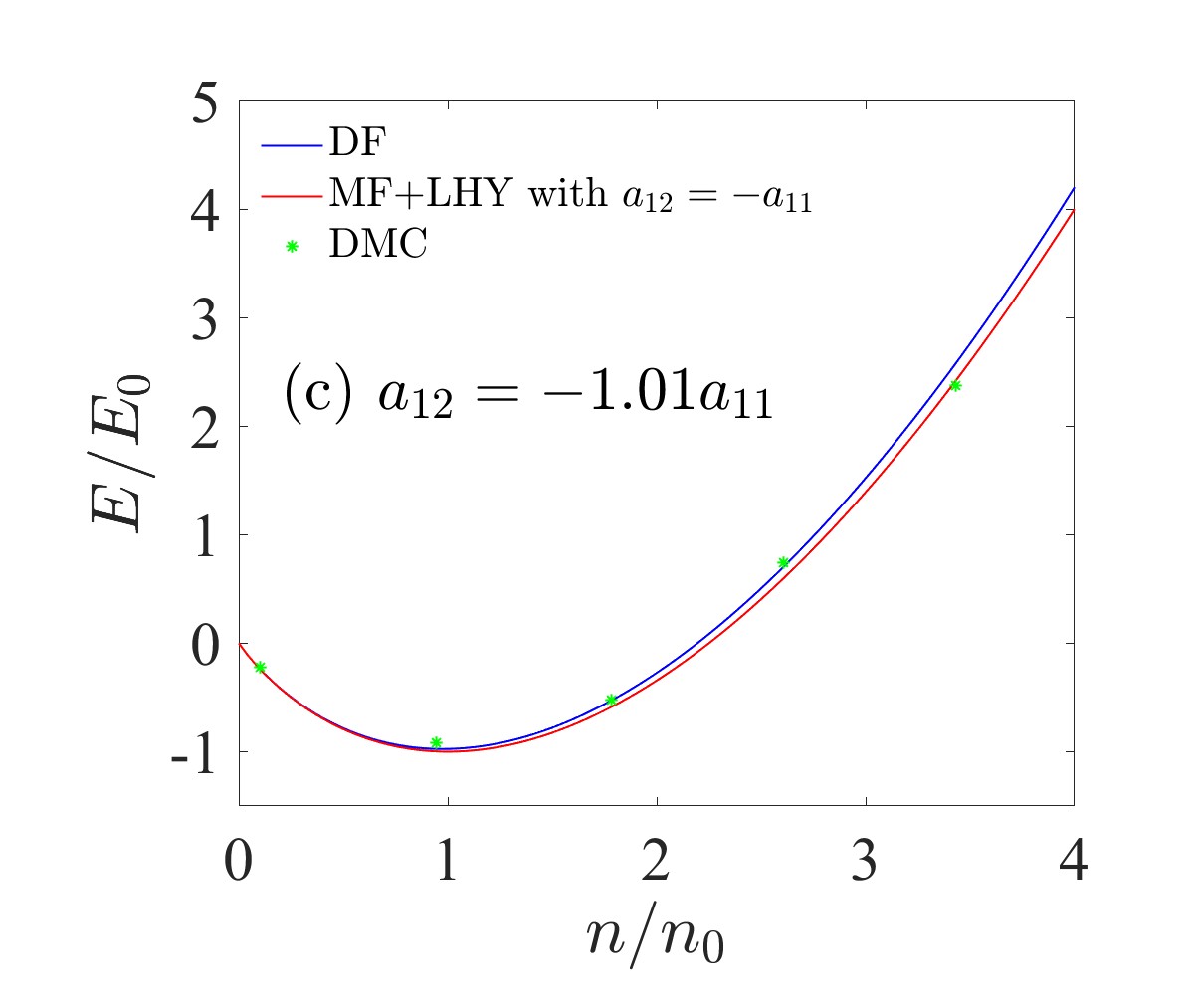}
		\caption{For the binary boson mixture droplet\cite{cikojevic2019universality}, equations of state  as predicted by our density-functional theory, DMC and
			the MF+LHY with $\abs{a_{12}}=a_{11}$\cite{Petrov2015}, for different values of the interspecies scattering length $a_{12}$. The green dots are the DMC results\cite{cikojevic2019universality}. The blue line shows our results from Eq.~(\ref{eos1}), and the red line shows the MF+LHY prediction with $\abs{a_{12}}=a_{11}$ as done in Ref. \cite{Petrov2015}. The definitions of energy units $E_0$ and density units $n_0$ are consistent with Ref. \cite{cikojevic2019universality}.}	\label{eos}
	\end{figure*}

	From Eq. (\ref{ec1}), the equation of state for 
	$g_{22} = g_{11}$ is given by
	\begin{gather}\label{eos1}
		\frac{E}{N}=\frac{(\mathcal{E}_{MF}+\mathcal{E}_C)V}{N}\\ \notag
		=\frac{\hbar^2\pi(a_{11}+a_{12})n}{m}+\frac{32\sqrt{2\pi}\hbar^2a_{11}^{5/2}}{15m}(1-\frac{a_{12}}{a_{11}})^{\frac{5}{2}}n^{\frac{3}{2}}, 
	\end{gather}
	where $a_{\sigma\sigma'}=\frac{mg_{\sigma\sigma'}}{4\pi\hbar^2}$ is the $s$-wave scattering length and $n=N/V$ is the density of each component. In Fig~.\ref{eos}, we show our results from Eq. (\ref{eos1}) for different values of the interspecies scattering length $a_{12}$, and compare them to the DMC equations of state\cite{cikojevic2019universality} and the MF+LHY prediction\cite{Petrov2015}. In Ref. \cite{Petrov2015}, to avoid the existence of imaginary parts of the LHY energy, $\abs{a_{12}}$ is approximated as $a_{11}$. The density-functional equations of state from Eq. (\ref{eos1}) do not contain imaginary parts and are closer to the DMC result than the MF+LHY prediction in Ref. \cite{Petrov2015} as $\frac{|a_{12}|}{a_{11}}$ increases, especially in the region with density less than the equilibrium density.  This is due to the fact that the quantum fluctuations are now treated self-consistently, better than the treatment in the Bogoliubov theory. The deviations from the DMC result at higher densities are probably due to higher-order effects neglected in our approach.
	
	\subsubsection{Dipolar Bose gas}
For a uniform dipolar Bose gas with all the dipoles aligned in $z$-direction, the renormalized Bogoliubov Hamiltonian corresponding to the quadratic action $S_2$ is given by 
\begin{gather}
	H_{Eff} \\ \notag
	=\sum_{\vb k} \epsilon_{k} a^\dagger_{\vb k} a_{\vb k}+\sum_{\vb k}U'({\vb k}) n [a^\dagger_{\vb k} a_{\vb k}+\frac{1}{2} ( a_{\vb k} a_{-\vb k}+H. C. ) ],
\end{gather}
where $U'({\vb k})=U({\vb k})+\chi$, $U({\vb k})=g[1+\epsilon_{dd}(3\cos^2 \phi_{\vb k}-1)]$, $g$ is the s-wave coupling constant, $\epsilon_{dd}$ is the strength of the dipole-dipole interaction, and $\phi_{\vb k}$ is the angle between $\vb k$ and the $z$-axis. The correlation energy
is given by \cite{lima2012}
\begin{equation}\label{dec}
	\mathcal{E}_C=\frac{64}{15\sqrt{\pi}}g'n^2 \sqrt{n a'^3}Q_5(\epsilon'_{dd}),
\end{equation}
where $g'=g+\chi$, $\epsilon'_{dd}=g \epsilon_{dd}/g'$, $a'=m g'/(4\pi\hbar^2)$,  $$Q_5(x)=\frac{(3x)^{5/2}}{48}[(8+26y+33y^2)\sqrt{1+y}$$
$$+15y^3\ln\frac{1+\sqrt{1+y}}{\sqrt{y}}],$$ and $y=(1-x)/3x.$
$\chi$ as a function of density can be solved self-consistently from Eq. (\ref{dec}) and (\ref{chi}). 
In the dilute limit, $g'\approx g$, the correlation energy is given by the LHY energy,
\begin{equation}\label{ddec}
	\mathcal{E}_C \approx \frac{64}{15\sqrt{\pi}}gn^2 \sqrt{n a^3}Q_5(\epsilon_{dd}),
\end{equation}
where $a= m g/(4\pi\hbar^2)$.  The effective correction to the $s$-wave coupling constant is approximately given by
\begin{equation}
	\chi \approx \frac{16}{\sqrt{\pi}}g \sqrt{n a^3}Q_5(\epsilon_{dd}),
\end{equation}
as found in Ref. \cite{zhang2022phonon}. 

\begin{figure}[htbp] 
	\centering 
	\includegraphics[width=0.4\textwidth]{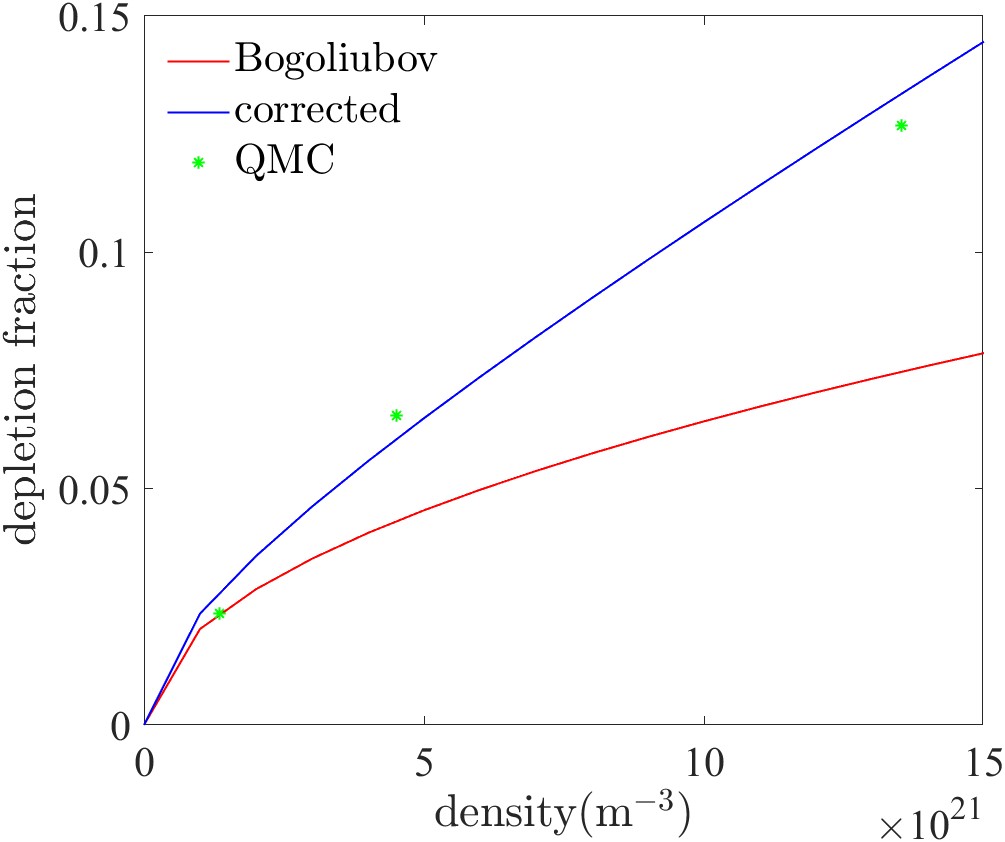} 
	\caption{For the $^{162}\rm{Dy}$ droplet\cite{bottcher2019dilute}, depletion fraction as predicted by our density-functional theory, QMC and
		the Bogoliubov theory, for a
		scattering length of $a = 60a_0$. The green dots are the QMC results\cite{bottcher2019dilute}. The blue line shows our results from Eq.~(\ref{fdc}), and the red line shows the Bogoliubov theory results from  Eq.~(\ref{fdb}). } 
	\label{qdf} 
\end{figure}

For the quantum droplet, $\epsilon_{dd}>1$, the function $Q_5(\epsilon_{dd})$ has a small imaginary part, which is neglected in the EGPE\cite{wachtler2016ground,baillie2016self}.  As found in Ref. \cite{zhang2022phonon}, the renormalized parameter $\epsilon'_{dd}$ is less than one, and this imaginary-energy problem is artificial.  Here we adopt an improved approximation scheme for the dilute limit to avoid the imaginary-energy problem from the start.  The correlation energy from Eq. (\ref{dec}) is not approximated by Eq. (\ref{ddec}), but given by
\begin{equation}
	\mathcal{E}_C \approx \frac{64}{15\sqrt{\pi}}gn^2 \sqrt{n a^3}Q_5(\epsilon'_{dd}),
\end{equation}
where the density dependence of $Q_5(\epsilon'_{dd})$ is much weaker than the prefactors.  Thus the correction to the $s$-wave coupling constant is approximately given by
\begin{equation}
	\chi \approx \frac{16}{\sqrt{\pi}}g \sqrt{n a^3}Q_5(\epsilon'_{dd}),
\end{equation}
and the renormalized $s$-wave coupling constant $g'$ can be obtained self-consistently
\begin{equation}
	g'=g+ \frac{16}{\sqrt{\pi}}g\sqrt{n a^3}Q_5(\epsilon'_{dd}).
\end{equation}
Here we check the accuracy of this approximation by computing the quantum depletion fraction and comparing it with the Monte Carle result \cite{bottcher2019dilute}. For a uniform dipolar Bose gas with density $n$, in Bogoliubov theory, the quantum depletion fraction is given by\cite{lima2012}
\begin{equation}\label{fdb}
	f_d^B=\frac{8}{3}\sqrt{n a^3/\pi}Q_3(\epsilon_{{dd}}),
\end{equation}
where $$Q_3(x)=\frac{(3x)^{3/2}}{8} [(2+5y)\sqrt{1+y}+3y^2\ln\frac{1+\sqrt{1+y}}{\sqrt{y}}],$$$$ y=\frac{(1-x)}{3x}.$$
Using the renormalized $s$-wave coupling constant $g'$, we can obtain the corrected quantum depletion fraction given by
\begin{equation}\label{fdc}
	f_d^c=\frac{8}{3}\sqrt{n a'^3/\pi}Q_3(\epsilon'_{{dd}}).
\end{equation}
In Fig~.\ref{qdf} we show the comparison between the corrected depletion fraction $f^c_d$ with the quantum Monte
Carlo (QMC) calculation \cite{bottcher2019dilute} and Bogoliubov theory. Our results agree well with the QMC, especially in the interval with a larger quantum depletion fraction, and both deviate from the Bogoliubov theory as the density increases. This suggests that the dipolar quantum droplet have stronger quantum fluctuations than the Gaussian fluctuations described by the Bogoliubov theory.

\begin{figure}[H] 
	\centering 
	\includegraphics[width=0.4\textwidth]{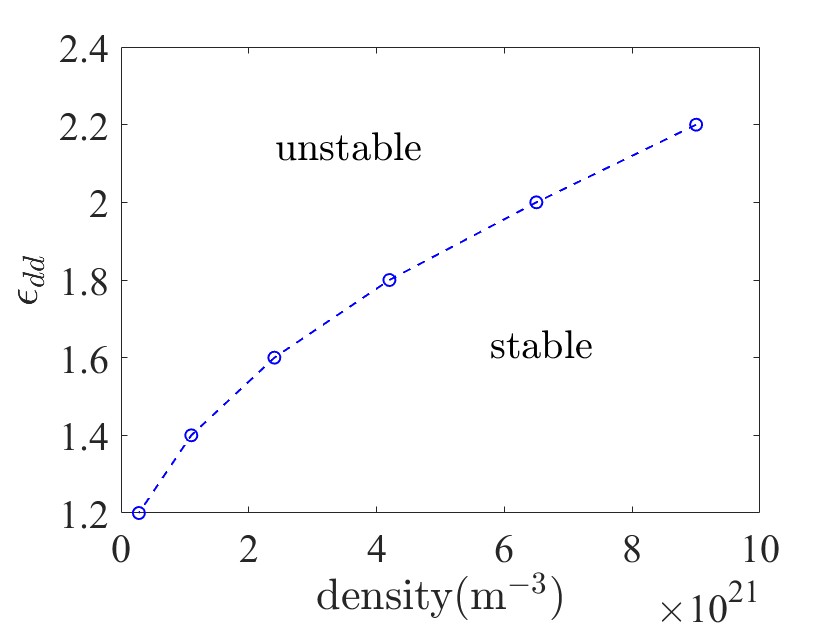} 
	\caption{For the $^{162}\rm{Dy}$ droplet\cite{bottcher2019dilute}, the stable region of the excitation spectrum as predicted by Eq. (\ref{rbes}). The blue line shows the stable boundary of the excitation spectrum.}  
	\label{eddn} 
\end{figure}

For dipolar Bose gases, the excitation energy in the Bogoliubov theory is given by
\begin{equation}
	\epsilon_B=\sqrt{\epsilon_k(2nU(\vb k)+\epsilon_k)}.
\end{equation}
In the quantum-droplet region with the strength of the dipole-dipole interaction $\epsilon_{{dd}}>1$, there is an imaginary part in the excitation energy for $ \phi_{\vb k}=\pi/2$, implying dynamical instability.  In our density-functional theory, the renormalized strength of the dipole-dipole interaction  $\epsilon'_{{dd}}$ can still be less than 1, thus stabilizing the excitation spectrum, which is given by
\begin{equation}
	\epsilon'_B=\sqrt{\epsilon_k(2nU'(\vb k)+\epsilon_k)}.\label{rbes}
\end{equation}
In Fig~.\ref{eddn}, we show the stable region of the renormalized excitation spectrum for different droplet densities and $\epsilon_{{dd}}$.
	\section{Discussion}
	It is worth mentioning that the density-functional theory was used to study $^4\rm{He}$ droplets \cite{casas1995density} where the interation parameters were treated phemenologically.  In the density-functional theory of quantum droplets, the quantum fluctuations renormalize the s-wave coupling constants, which can be determined self-consistently in the dilute region. For these experimental systems, the results from our approach are consistent with the EGPE but do not suffer from the imaginary-energy problem.  Our results for the ground-state energy and the quantum depletion fraction are in good agreement with the Monte Carlo results.  For systems with significant quantum depletion, our approach should be better as it treats quantum fluctuations self-consistently.    
	\section{acknowledgments}We would like to thank Z.-Q. Yu for helpful discussions.

	\bibliographystyle{unsrt}
	\bibliography{DFTQDNew}
	
\end{document}